\documentstyle[11pt,psfig,twoside]{article}

\begin{document}

\newcommand{\dd}{\,{\rm d}}
\newcommand{\ie}{{\it i.e.},\,}
\newcommand{\etal}{{\it et al.\ }}
\newcommand{\eg}{{\it e.g.},\,}
\newcommand{\cf}{{\it cf.\ }}
\newcommand{\vs}{{\it vs.\ }}
\newcommand{\zdot}{\makebox[0pt][l]{.}}
\newcommand{\up}[1]{\ifmmode^{\rm #1}\else$^{\rm #1}$\fi}
\newcommand{\dn}[1]{\ifmmode_{\rm #1}\else$_{\rm #1}$\fi}
\newcommand{\upd}{\up{d}}
\newcommand{\uph}{\up{h}}
\newcommand{\upm}{\up{m}}
\newcommand{\ups}{\up{s}}
\newcommand{\arcd}{\ifmmode^{\circ}\else$^{\circ}$\fi}
\newcommand{\arcm}{\ifmmode{'}\else$'$\fi}
\newcommand{\arcs}{\ifmmode{''}\else$''$\fi}
\newcommand{\MS}{{\rm M}\ifmmode_{\odot}\else$_{\odot}$\fi}
\newcommand{\RS}{{\rm R}\ifmmode_{\odot}\else$_{\odot}$\fi}
\newcommand{\LS}{{\rm L}\ifmmode_{\odot}\else$_{\odot}$\fi}

\newcommand{\Abstract}[2]{{\footnotesize\begin{center}ABSTRACT\end{center}
\vspace{1mm}\par#1\par
\noindent
{~}{\it #2}}}

\newcommand{\TabCap}[2]{\begin{center}\parbox[t]{#1}{\begin{center}
  \small {\spaceskip 2pt plus 1pt minus 1pt T a b l e}
  \refstepcounter{table}\thetable \\[2mm]
  \footnotesize #2 \end{center}}\end{center}}

\newcommand{\TableSep}[2]{\begin{table}[p]\vspace{#1}
\TabCap{#2}\end{table}}

\newcommand{\FigCap}[1]{\footnotesize\par\noindent Fig.\  %
  \refstepcounter{figure}\thefigure. #1\par}

\newcommand{\TableFont}{\footnotesize}
\newcommand{\TableFontIt}{\ttit}
\newcommand{\SetTableFont}[1]{\renewcommand{\TableFont}{#1}}

\newcommand{\MakeTable}[4]{\begin{table}[htb]\TabCap{#2}{#3}
  \begin{center} \TableFont \begin{tabular}{#1} #4 
  \end{tabular}\end{center}\end{table}}

\newcommand{\MakeTableSep}[4]{\begin{table}[p]\TabCap{#2}{#3}
  \begin{center} \TableFont \begin{tabular}{#1} #4 
  \end{tabular}\end{center}\end{table}}

\newenvironment{references}%
{
\footnotesize \frenchspacing
\renewcommand{\thesection}{}
\renewcommand{\in}{{\rm in }}
\renewcommand{\AA}{Astron.\ Astrophys.}
\newcommand{\AAS}{Astron.~Astrophys.~Suppl.~Ser.}
\newcommand{\ApJ}{Astrophys.\ J.}
\newcommand{\ApJS}{Astrophys.\ J.~Suppl.~Ser.}
\newcommand{\ApJL}{Astrophys.\ J.~Letters}
\newcommand{\AJ}{Astron.\ J.}
\newcommand{\IBVS}{IBVS}
\newcommand{\PASP}{P.A.S.P.}
\newcommand{\Acta}{Acta Astron.}
\newcommand{\MNRAS}{MNRAS}
\renewcommand{\and}{{\rm and }}
\section{{\rm REFERENCES}}
\sloppy \hyphenpenalty10000
\begin{list}{}{\leftmargin1cm\listparindent-1cm
\itemindent\listparindent\parsep0pt\itemsep0pt}}%
{\end{list}\vspace{2mm}}

\def\TYLDA{~}
\newlength{\DW}
\settowidth{\DW}{0}
\newcommand{\dw}{\hspace{\DW}}

\newcommand{\refitem}[5]{\item[]{#1} #2%
\def\REFARG{#3}\ifx\REFARG\TYLDA\else, {\it#3}\fi
\def\REFARG{#4}\ifx\REFARG\TYLDA\else, {\bf#4}\fi
\def\REFARG{#5}\ifx\REFARG\TYLDA\else, {#5}\fi.}

\newcommand{\Section}[1]{\section{#1}}
\newcommand{\Subsection}[1]{\subsection{#1}}
\newcommand{\Acknow}[1]{\par\vspace{5mm}{\bf Acknowledgements.} #1}
\pagestyle{myheadings}

\def\thefootnote{\fnsymbol{footnote}}
\begin{center}

{\Large\bf  The Optical Gravitational Lensing Experiment.\\
\vskip3pt
Multiple  Cluster Candidates in the Small\\
\vskip3pt
Magellanic Cloud\footnote{Based on  observations obtained with the 1.3~m
Warsaw telescope at the Las Campanas  Observatory of the Carnegie
Institution of Washington.}}

\vskip 1cm

{G.~~P~i~e~t~r~z~y~\'n~s~k~i,~~ and~~A.~~U~d~a~l~s~k~i}
\vskip5mm
{Warsaw  University Observatory, Al.~Ujazdowskie~4, 00-478~Warszawa,
Poland\\  e-mail: (pietrzyn,udalski)@sirius.astrouw.edu.pl} 

\end{center}

\Abstract{We present the list of potential multiple star clusters from the 
central part of the SMC. Presented systems were selected from the catalog of 
star clusters from the SMC. We find 23 suspected cluster pairs and 4 triple 
systems. The statistical analysis suggests that many of them may constitute 
physical systems. Size, equatorial coordinates and age of presented clusters 
are given. Age of clusters which form five pairs and one triple system is 
coeval suggesting common origin of these objects.}{~}

\Section{Introduction}
The presence of large number of potential  pairs of clusters in the Magellanic 
Clouds is well established. The catalog of binary star clusters from the SMC  
was presented by Hatzidimitriou and Bhatia (1990). The possible cluster pairs  
from the LMC were cataloged by Bhatia and Hatzidimitriou (1988) and  Bhatia 
\etal (1991). Statistical analysis made by these authors suggests that many 
of them may be  real binary systems. Several recent papers deal with  selected 
binary pairs from the LMC. Based on spectroscopic and photometric  
observations of three pairs Kontizas \etal (1993) concluded that all three are 
very young physical systems. Dieball and Grebel (1998) claimed physical nature 
of the pair SL~538 and NGC~2006. Surface brightness photometry made for three 
pairs by Vallenari \etal (1998) indicates interaction between physically 
connected clusters. Leon \etal (1998) pointed out that several cluster pairs  
show tidal tails, thus they are real binary clusters. All mentioned studies 
confirmed the fact that significant fraction of cataloged binary cluster 
candidates are indeed physical pairs. 

Although the pairs from the LMC were subject of several investigations in the 
near past, to our knowledge no papers on pairs from the SMC were published 
since the catalog of Hatzidimitriou and Bhatia (1990). 

The existence of physical pairs of clusters has important implication on the 
process of formation and evolution of clusters. Fujimoto and Kumai (1997) 
suggest that binary and multiple clusters form through oblique collisions 
between massive gas clouds. This mechanism leads to the cluster systems having 
similar age. The observations of cluster pairs that show large  age difference 
between components (Vallenari \etal 1998) cannot be, however, explained by 
this theory.  Leon \etal (1999) proposed scenario of tidal capture in the 
groups  of clusters for explaining such systems. 

The main purpose of the OGLE microlensing project is detection of dark matter in the Galaxy 
with microlensing phenomena (Paczy{\'n}ski 1986). Because of very low 
probability of detection of microlensing events the project requires monitoring 
of million stars. The observations are conducted in the Galactic bulge,  
Galactic disc and central regions of the Magellanic Clouds. The OGLE-II phase 
of the project is described by Udalski, Kubiak and Szyma{\'n}ski (1997). Huge 
amount of precise photometric data for stars from very dense regions, rarely 
observed until now with the modern instruments, provides an unique material 
for many other studies. In particular data collected for the regions located 
in the Magellanic Clouds are very well suited for studying the rich system of 
clusters from these galaxies. The observational material obtained for the SMC 
was presented by Udalski \etal (1998). It contains {\it BVI} photometric and 
astrometric data for more than 2 million stars from the 2.4 square degree 
region located in the center of the SMC. Based on these data Pietrzy{\'n}ski 
\etal (1998) presented  the catalog of 238  star clusters. In this paper we 
use this catalog to identify possible binary and multiple star clusters. 

\Section{Cluster Pairs}

We searched the OGLE catalog of clusters from the SMC (Pietrzy{\'n}ski \etal 
1998) for objects with the  projected separations smaller than 18~pc 
(Hatzidimitriou and Bhatia 1990), assuming the distance to the SMC of 54~kpc 
(Udalski 1998). Based on the positions of 161 clusters detected with the 
algorithmic, automatic method, eight pairs and three triple systems  were 
detected. Table~1 contains their description. However, the OGLE catalog of 
clusters also contains data for additional 77 clusters found during visual 
examination of observed regions. They were too small and faint to be detected 
with the automatic procedure. If we include their coordinates to our search we 
find additional 15 pairs and one triple system. Since these objects are less 
reliable we present them separately in Table~2.  Cluster coordinates and sizes 
are extracted from the OGLE catalog of clusters (Pietrzy{\'n}ski \etal 1998). 
Age is taken from Pietrzy{\'n}ski and Udalski (1999). 
\MakeTable{|c|c|c|c|c|}{10cm}{Multiple cluster candidates in the SMC}{
\hline
Name  & $\alpha_{2000}$ & $\delta_{2000}$ & Radius & $\log t$\\
OGLE-CL-& &  & [\arcs] & \\ \hline
SMC0018 & $0\uph43\upm37\zdot\ups57$ & $-73\arcd26\arcm37\zdot\arcs9$ & 20 & 7.9\\
SMC0017 & $0\uph43\upm32\zdot\ups74$ & $-73\arcd26\arcm25\zdot\arcs4$ & 26 & 7.9\\ \hline
SMC0021 & $0\uph43\upm44\zdot\ups40$ & $-72\arcd58\arcm35\zdot\arcs6$ & 36 & -- \\
SMC0020 & $0\uph43\upm37\zdot\ups89$ & $-72\arcd58\arcm48\zdot\arcs3$ & 9  & 8.6 \\
SMC0019 & $0\uph43\upm37\zdot\ups59$ & $-72\arcd57\arcm30\zdot\arcs9$ & 12 & 8.6 \\ \hline
SMC0042 & $0\uph47\upm49\zdot\ups72$ & $-73\arcd28\arcm42\zdot\arcs2$ & 16 & --  \\
SMC0045 & $0\uph48\upm00\zdot\ups68$ & $-73\arcd29\arcm10\zdot\arcs3$ & 35 & -- \\ \hline
SMC0063 & $0\uph50\upm36\zdot\ups83$ & $-73\arcd03\arcm28\zdot\arcs0$ & 30 &  -- \\
SMC0065 & $0\uph50\upm54\zdot\ups62$ & $-73\arcd03\arcm26\zdot\arcs9$ & 20 &  -- \\ \hline
SMC0077 & $0\uph52\upm13\zdot\ups34$ & $-73\arcd00\arcm12\zdot\arcs2$ & 18 & 7.9 \\
SMC0078 & $0\uph52\upm16\zdot\ups56$ & $-73\arcd01\arcm04\zdot\arcs0$ & 36 & 7.9\\ \hline
SMC0084 & $0\uph52\upm46\zdot\ups69$ & $-73\arcd24\arcm25\zdot\arcs4$ & 12 & -- \\
SMC0087 & $0\uph52\upm48\zdot\ups99$ & $-73\arcd24\arcm43\zdot\arcs3$ & 22 & 8.7 \\ \hline
SMC0093 & $0\uph53\upm31\zdot\ups29$ & $-72\arcd40\arcm04\zdot\arcs2$ & 18 & -- \\
SMC0094 & $0\uph53\upm40\zdot\ups09$ & $-72\arcd39\arcm35\zdot\arcs3$ & 9  & -- \\
SMC0096 & $0\uph53\upm42\zdot\ups31$ & $-72\arcd39\arcm14\zdot\arcs6$ & 11 & -- \\ \hline
SMC0114 & $0\uph58\upm25\zdot\ups73$ & $-72\arcd39\arcm56\zdot\arcs5$ & 18 & -- \\
SMC0113 & $0\uph58\upm16\zdot\ups29$ & $-72\arcd38\arcm46\zdot\arcs8$ & 24 & -- \\ \hline
SMC0119 & $0\uph59\upm56\zdot\ups87$ & $-72\arcd22\arcm24\zdot\arcs4$ & 9  & -- \\
SMC0120 & $1\uph00\upm01\zdot\ups33$ & $-72\arcd22\arcm08\zdot\arcs7$ & 27 & 7.7 \\ \hline
SMC0138 & $1\uph03\upm53\zdot\ups02$ & $-72\arcd06\arcm10\zdot\arcs5$ & 18 &  7.4 \\
SMC0144 & $1\uph04\upm05\zdot\ups23$ & $-72\arcd07\arcm14\zdot\arcs6$ & 18 &  7.6 \\ \hline
SMC0146 & $1\uph05\upm13\zdot\ups40$ & $-71\arcd59\arcm41\zdot\arcs8$ & 14 & 7.3 \\
SMC0145 & $1\uph05\upm04\zdot\ups30$ & $-71\arcd59\arcm24\zdot\arcs8$ & 18 & 7.9 \\
SMC0147 & $1\uph05\upm07\zdot\ups95$ & $-71\arcd59\arcm45\zdot\arcs1$ & 22 & 7.1 \\ \hline
}

In order to check whether the number of detected pairs is significantly 
different than the number expected from chance line-up due to projection we 
conducted statistical analysis described by Bhatia and Hatzidimitriou (1988). 
The number of chance-pairs of objects uniformly distributed in the space may 
be calculated based on the formula given by Page (1972). 
$$N_{1}=0.5\pi\times N_{2}^{2}\times s^{2}$$
where $N_{1}, N_{2}$ and $s$ are the  expected number of pairs per square 
degree, the number of clusters per square degree and projected angular 
separation in degrees, respectively. The OGLE catalog contains 238 clusters 
located in the 2.4 square degree region of the SMC. Assuming that they are 
distributed uniformly we find that the expected number of chance-pairs with 
separation smaller than 18 pc should be about eight (four in the case of 161 
clusters detected in the algorithmic way). In fact the distribution of 
clusters is not uniform. We account for that adopting this procedure to  
${15\times15}$~arcmin regions for which one may assume uniform density of 
clusters. We obtain that the number of chance-pairs is about ten and seven 
using positions of all clusters and 161 detected automatically, respectively. 
As one can notice these numbers are significantly smaller than the number of 
detected cluster pairs. If we assume Poissonian statistics we obtain that the 
difference is about $7\sigma$ which suggests that the majority of presented 
pairs constitute physical systems.

From Tables~1 and 2 one can see that five pairs and one triple system have 
nearly the same age. This fact favors the common origin of these systems. 
Their sizes are also comparable. 

\Section{Summary}
We present the list of potential binary and multiple clusters from the center 
of the SMC. Altogether 23 pairs and 4 triple systems are selected. Based on 
statistical considerations we show that the number of expected chance-pairs of 
clusters is considerably smaller. This fact suggests that most of them must 
constitute real physical systems. For 22 members of detected pairs their age 
was determined. Most of them are found to be young objects. Members of five 
pairs  and one triple system have similar age which indicates their common 
origin. Further observations, in particular spectroscopic, are required to 
confirm physical relation of presented systems. 

\renewcommand{\arraystretch}{1.05}
\MakeTable{|c|c|c|c|c|}{10cm}{Additional multiple cluster candidates}{
\hline
Name  & $\alpha_{2000}$ & $\delta_{2000}$ & Radius & $\log t$\\
OGLE-CL- & &  & [\arcs] & \\ \hline
SMC0005 &  $0\uph39\upm21\zdot\ups63$ &  $-73\arcd15\arcm28\zdot\arcs4$ &  22 & -- \\
SMC0165 &  $0\uph39\upm11\zdot\ups56$ &  $-73\arcd14\arcm45\zdot\arcs5$ &  12 & -- \\ \hline
SMC0033 &  $0\uph46\upm12\zdot\ups26$ &  $-73\arcd23\arcm34\zdot\arcs0$ &  18 & 7.2 \\
SMC0182 &  $0\uph46\upm01\zdot\ups63$ &  $-73\arcd23\arcm44\zdot\arcs4$ &  7 & -- \\ \hline
SMC0035 &  $0\uph46\upm33\zdot\ups72$ &  $-72\arcd46\arcm25\zdot\arcs9$ &  14 & -- \\
SMC0185 &  $0\uph46\upm34\zdot\ups04$ &  $-72\arcd45\arcm55\zdot\arcs7$ &  4 & -- \\ \hline
SMC0040 &  $0\uph47\upm01\zdot\ups18$ &  $-73\arcd23\arcm34\zdot\arcs7$ &  16 & -- \\
SMC0187 &  $0\uph47\upm05\zdot\ups87$ &  $-73\arcd22\arcm16\zdot\arcs6$ &  14 & -- \\ \hline

SMC0190 &  $0\uph48\upm13\zdot\ups20$ &  $-72\arcd47\arcm34\zdot\arcs7$ &  12 & --\\
SMC0191 &  $0\uph48\upm20\zdot\ups19$ &  $-72\arcd47\arcm42\zdot\arcs1$ &  20 & --\\ \hline
SMC0060 &  $0\uph50\upm21\zdot\ups95$ &  $-73\arcd23\arcm16\zdot\arcs5$ &  36 & -- \\
SMC0197 &  $0\uph50\upm03\zdot\ups82$ &  $-73\arcd23\arcm03\zdot\arcs9$ &  24 & 8.4 \\ \hline
SMC0059 &  $0\uph50\upm16\zdot\ups06$ &  $-73\arcd01\arcm59\zdot\arcs6$ &  25 & 7.8 \\
SMC0199 &  $0\uph50\upm15\zdot\ups07$ &  $-73\arcd03\arcm14\zdot\arcs6$ &   7 & -- \\ \hline
SMC0064 &  $0\uph50\upm39\zdot\ups55$ &  $-72\arcd57\arcm54\zdot\arcs8$ &  36 & 8.1\\
SMC0200 &  $0\uph50\upm38\zdot\ups98$ &  $-72\arcd58\arcm43\zdot\arcs6$ &  11 & 8.0 \\ \hline
SMC0210 &  $0\uph52\upm30\zdot\ups30$ &  $-73\arcd02\arcm59\zdot\arcs0$ &  21 & 8.2 \\
SMC0211 &  $0\uph52\upm32\zdot\ups15$ &  $-73\arcd02\arcm10\zdot\arcs3$ &  14 & -- \\ \hline
SMC0083 &  $0\uph52\upm44\zdot\ups27$ &  $-72\arcd58\arcm47\zdot\arcs8$ &  29 & 7.8 \\
SMC0212 &  $0\uph52\upm44\zdot\ups52$ &  $-72\arcd59\arcm24\zdot\arcs2$ &   9 & --\\
SMC0213 &  $0\uph52\upm48\zdot\ups29$ &  $-72\arcd59\arcm22\zdot\arcs2$ &  11 & -- \\ \hline
SMC0097 &  $0\uph54\upm11\zdot\ups00$ &  $-72\arcd51\arcm54\zdot\arcs1$ &  20 & -- \\
SMC0217 &  $0\uph53\upm56\zdot\ups49$ &  $-72\arcd51\arcm23\zdot\arcs8$ &   8 & -- \\ \hline
SMC0108 &  $0\uph56\upm34\zdot\ups48$ &  $-72\arcd30\arcm08\zdot\arcs3$ &  15 & -- \\
SMC0223 &  $0\uph56\upm25\zdot\ups59$ &  $-72\arcd29\arcm45\zdot\arcs1$ &   6 & -- \\ \hline
SMC0112 &  $0\uph57\upm57\zdot\ups14$ &  $-72\arcd26\arcm42\zdot\arcs0$ &  29 & 7.5 \\
SMC0227 &  $0\uph57\upm50\zdot\ups23$ &  $-72\arcd26\arcm23\zdot\arcs6$ &   7 & -- \\ \hline
SMC0123 &  $1\uph00\upm33\zdot\ups09$ &  $-72\arcd14\arcm23\zdot\arcs0$ &  31 & -- \\ 
SMC0230 &  $1\uph00\upm33\zdot\ups15$ &  $-72\arcd15\arcm30\zdot\arcs5$ &   9 & -- \\ \hline
SMC0231 &  $1\uph00\upm58\zdot\ups19$ &  $-72\arcd32\arcm24\zdot\arcs9$ &  21 & -- \\
SMC0232 &  $1\uph01\upm13\zdot\ups58$ &  $-72\arcd33\arcm03\zdot\arcs5$ &   8 & -- \\ \hline
SMC0139 &  $1\uph03\upm53\zdot\ups44$ &  $-72\arcd49\arcm34\zdot\arcs2$ &  20 & 7.5 \\
SMC0235 &  $1\uph03\upm58\zdot\ups96$ &  $-72\arcd48\arcm18\zdot\arcs3$ &   8 & -- \\ \hline
}

Tables~1 and 2 are available from the OGLE Internet archive: \newline
{\it http://www.astrouw.edu.pl/\~{}ftp/ogle}. Photometric data and finding
charts for all clusters can also  be found there. 

\Acknow{The paper was partly supported by the Polish KBN grant 2P03D00814 to 
A.~Udalski.  Partial support for the OGLE project was provided with the NSF 
grant  AST-9530478 to B.~Paczy\'nski.}

\end{document}